\documentclass[%
 reprint,
amsmath,
amphysicalsymb,
aps,
pra,
nofootinbib
]{revtex4-1}

\usepackage{graphicx}
\usepackage{dcolumn}
\usepackage{bm}
\usepackage[utf8]{inputenc} 
\usepackage[T1]{fontenc} 
\usepackage{hyperref}
\usepackage{url} 
\usepackage{booktabs}  
\usepackage{amsfonts} 
\usepackage{microtype} 
\usepackage{amssymb,amsmath,amsthm,graphicx}
\usepackage{xcolor}
\usepackage{comment}
\usepackage{multirow}
\usepackage[caption=false]{subfig}
\usepackage{tabularx}
\usepackage{float}
\usepackage{algorithm}
\usepackage{algpseudocode}
\usepackage{booktabs}
\usepackage{enumitem} 
\usepackage{lipsum}
\usepackage{physics}
\usepackage{bbm}
\usepackage{todonotes}
\algrenewcommand\algorithmicrequire{\textbf{Input:}}
\algrenewcommand\algorithmicensure{\textbf{Output:}}

\begin{document}

\title{Efficient Quantum Error Mitigation for Unitary $k$-Designs}

\author{Ayush Pancholy}
\email{ayush.pancholy@berkeley.edu}

\author{K. Birgitta Whaley}
\affiliation{Berkeley Center for Quantum Information and Computation, Berkeley, California 94720, USA}
\affiliation{Department of Chemistry, University of California, Berkeley, California 94720, USA}

\date{\today}
\begin{abstract}
    Quantum circuit ensembles that have the properties of unitary $k$-designs represent applications where there is no obvious bias toward any particular Pauli support, as is the case in simulating systems exhibiting ``quantum chaos,'' which range from quantum dynamics near black holes to gapless spin fluid analysis. However, noisy hardware makes quantum circuits prone to a myriad of error sources, of which depolarizing and coherent error can be particularly destructive. To combat depolarizing error, popular techniques typically involve circuit or gate folding, which can be time-intensive procedures due to increased circuit depth and shot overhead.  Other tensor-network-based mitigation techniques suffer from intractability in high-entanglement regimes. In this work, we leverage the structure of unitary $k$-design Pauli support distributions by introducing a technique we name ``circuit balancing,'' along with gate benchmarking data, in order to estimate circuit-wide depolarization. We describe how to invert the diagnosed circuit depolarization even in the presence of coherent error, via Pauli twirling. We provide asymptotics to estimate the number of twirls needed to maintain a desired output fidelity. We test our method numerically in a variety of simulation settings and find that it can significantly reduce average random circuit infidelity. Further, we employ our methods to find significant infidelity reductions when running a random circuit ensemble on a contemporary superconducting quantum computer, IBM Fez. Overall, we show that the method effectively reduces gate-based error for unitary $k$-designs
    without incurring any two-qubit gate overhead.
\end{abstract}
\maketitle 

\section{Introduction}
In recent years, the task of running random circuits on quantum hardware has become a problem of considerable interest, and faithfully executing random circuit ensembles has emerged as a benchmarking test for the capability of quantum computers in the noisy intermediate scale quantum (NISQ) regime \cite{randomizedbenchmarking2011}, and when combined with randomized compiling, randomized benchmarks provide a pathway to test for fault tolerance in the context of error correction \cite{benchmarkingbelowthresh}. However, circuits that represent unitary $k$-designs are also associated with applications where there is no naturally expected bias toward a particular Pauli support. 
These types of applications are often referred to as exhibiting ``quantum chaos:'' examples include simulating quantum dynamics near black holes and gapless spin fluid analysis \cite{quantumchaos}. As such, circuit ensembles with limited distinguishability from the Haar distribution not only have benchmarking utility but also carry physically meaningful significance for simulation purposes. We also anticipate numerous other applications whose representative circuit ensembles are sufficiently deep and entangling so as to approximate unitary 2-designs. In this work we address the question of error mitigation schemes for circuits representing unitary $k$-designs and seek to suppress major contributions of device noise without significant quantum overhead.

Two major types of error on quantum devices are depolarizing noise channels and coherent error channels. Other types of error include idling rotations, state preparation and measurement (SPAM) error, and relaxation. However, it is well known that dominant idling errors can be corrected by appropriately placed dynamical decoupling sequences on many device architectures \cite{qctrldd2024} \cite{trappediondd}, while SPAM errors with moderate correlation can be suppressed by measurement error mitigation techniques, although with a shot overhead \cite{ibmmeasurementmitigation2021}. Under the assumption that relaxation effects are minimal for circuits below a specific depth, in this work we seek to address depolarizing and coherent error induced by noisy gates in scrambling circuits representing unitary $k$-designs.

Many existing techniques attempt to combat depolarizing noise by benchmarking the depolarizing parameter of circuits via repetition sequences, as in Zero-Noise Extrapolation (ZNE) \cite{xanaduzne2021} \cite{znebestpractices2023}. However, when running an ensemble of circuits, it is often intractable to run motifs of circuit folds since this significantly increases circuit depth. This increase in depth can incur relaxation effects, or simply add too much runtime for the target task. Other techniques, such as Probabilistic Error Cancellation (PEC), although successful at capturing complex error sources, can incur an even greater computational overhead than ZNE \cite{ibmpec2017} \cite{pecpaulilinblad2023}. 

Another class of error mitigation methods involves modeling the composition of device errors via tensor network simulation and inversion through classical postprocessing  \cite{tensornetworkmitigation2023} \cite{tensornetworknoisechar2024}. These are collectively known as Tensor-Network Error Mitigation (TEM) techniques. However, these methods suffer when system entanglement is no longer classically representable, limiting the regime in which they can be successfully applied. 

Error mitigation for unitary $k$-designs is often treated with the same methods as for general circuits (using ZNE, PEC, or TEM), and the thresholds for accurate noise inversion are known over the regime of random circuit sampling \cite{thresholdsfornoisyrqc2025}. In some cases, error mitigation has not even been applied 
to random circuit runs, notably in the landmark quantum supremacy experiment conducted by Google, which relied instead on hardware advancements and traditional error suppression techniques \cite{quantumsupremacy2019}.

The key novelty of the current work is to analytically leverage the structure of unitary $k$-design Pauli supports in order to estimate the depolarizing parameter of a target circuit via benchmarking data. This procedure requires an optimization-based layout selection technique we name ``circuit balancing,'' which is used to ensure that the local depolarization affecting each qubit pair is similar. We then invert the circuit depolarization from this diagnostic at the bitstring level and project the results onto the probability simplex so as to produce a valid distribution. We also leverage Pauli twirling to avoid the negative impacts of coherent error on circuit depolarization estimation.

The structure of the remainder of this paper is as follows. We first provide a background and summary of related work in Section~\ref{sec:background}. In Section \ref{sec:pauli_support}-\ref{sec:global_twoq_depolarization}, we then discuss why it is possible to estimate a global depolarizing parameter for a unitary $k$-design, together with necessary assumptions on circuit structure. In Section \ref{sec:circuit_balancing} we provide a layout selection technique that enables approximate satisfaction of these assumptions and Section~\ref{sec:global_parameter_estimation} then summarizes the parameter estimation protocol. In Section \ref{sec:inversion}, we discuss our approach to noise inversion, with construction of an asymptotic scaling estimate on the number of Pauli twirls needed to faithfully invert our depolarizing noise in the presence of coherent error (Section \ref{sec:twirl_overhead}). In Section \ref{sec:numerical_experiments}, we then provide simulation data illustrating the efficacy and robustness of our approach, and in Section \ref{sec:hardware_experiments}, we demonstrate our methods on real quantum hardware. Section \ref{sec:conclusions} provides a conclusion and outlook for future work.

\section{Background \& Related Work} \label{sec:background}
Error mitigation approaches that attempt to invert depolarizing noise have a long history. As mentioned, the original naive ZNE approach can be computationally costly. The protocol estimates the global depolarizing parameter of a circuit implementing $U$ by repeating motifs of $(U^\dagger U)^kU$ for varying $k$, called ``folding.'' Each addition of $U$ or $U^\dagger$ introduces another factor of gate and depth overhead. 
Additionally, repetitions of this unitary motif amplify noise sources like depolarization, but do not amplify coherent error. 
Similar approaches have been taken to fold at the gate level rather than at the unitary level, although these methods incur both gate and circuit cost as well while being insensitive to coherent error at the gate level \cite{xanaduzne2021}. 

Despite its success in reducing error on wide circuits run with noisy quantum hardware \cite{compreach2018} \cite{scalableerrormit2023}, the significant experimental overhead of ZNE has prompted systematic attempts to reduce computational cost. These include identity insertion approaches \cite{identityinsertions2020} \cite{invertedcircuits2024} in order to reduce gate counts in repetition motifs, and also advancements in the extrapolation mechanisms themselves, to improve error estimation without incurring significant additional overhead \cite{richardsonextrap2022} \cite{analysiszne2025}. 
Folding-free ZNE protocols have also been suggested that use circuit reliability metrics and extrapolate the noise regimes of subcircuits within the target circuits \cite{foldingfreezne2023}. Additionally, the idea of using hardware benchmarking data to improve the efficacy of folding techniques has been proposed \cite{noiseawarezne2024}. One variant of traditional ZNE that has been used with some success is Probabilistic Error Amplification (PEA), although this technique requires the overhead of learning the error rates of layers of gates in a particular circuit \cite{evidenceofutility2023}. 
The related Probabilistic Error Cancellation (PEC) approach can successfully capture complex, correlated errors, but at the expense of a generally significantly greater overhead compared to ZNE \cite{pecpaulilinblad2023}. 

Other techniques for error mitigation exist that successfully limit quantum overhead, such as the aforementioned TEM approaches. These methods encode the error propagation into a tensor network representing the quantum error channel, often using information about device-level error, to enable application of noise inversion in postprocessing \cite{tensornetworkmitigation2023} \cite{tensornetworknoisechar2024}. It is also possible to bound the noise characterization threshold needed to successfully run random quantum circuits on noisy devices with mitigation schemes like TEM and PEC \cite{thresholdsfornoisyrqc2025}. However, an implicit assumption of TEM-based approaches is that circuit entanglement is low enough for the inversion to be tractable. 
In this work, we explore a novel type of noise inversion that is dependent on gate-level benchmarking data. Our mitigation scheme differs from existing approaches by leveraging the properties of unitary $k$-designs run on devices with specific (but common) connectivity graphs, which allow us to efficiently estimate and invert depolarizing error solely from benchmarking data. Our approach also circumvents the large quantum overhead of folding-based techniques and has no preferred low-entanglement regime, unlike TEM.

Our work relies on data produced via regular device benchmarking. We expect hardware manufacturers to routinely benchmark device gates after calibrations. Assessing gate-level error, particularly depolarizing error, is typically done via some form of randomized benchmarking, including Cycle Benchmarking (CB) \cite{cyclebenchmarking2019}, Direct Randomized Benchmarking \cite{directrb} \cite{directrbtheory}, and Interleaved Randomized Benchmarking (iRB) \cite{irb2012}. All of these methods produce estimates of gate infidelity, and Direct RB in particular produces perfect estimates of average gate error when the sole source is depolarization. Statistical error can be arbitrarily suppressed by increasing the number of benchmarking sequences and shot count. Additionally, Direct RB is naturally insensitive to SPAM error due to its fitting technique, and does not require compiling arbitrary Clifford sequences into a target gateset, instead leveraging random circuit construction intrinsically from a desired gateset \cite{directrbtheory}. Finally, randomized benchmarking techniques have desirable noise tailoring properties, as the random sequences are similar to randomized compiling, which effectively converts coherent rotations to stochastic Pauli errors \cite{wallman2015}. As a result, periodic benchmarking using Direct RB-like techniques would yield routine access to high-fidelity estimates of gate-level depolarizing error, and the precision of the estimate can be controlled by increasing the shot or sequence overhead.

\section{Depolarizing Parameter Estimation \label{sec:parameter_estimation}}
\subsection{Pauli Support Distribution for Unitary $1$-Designs \label{sec:pauli_support}}
Consider a unitary $U \sim \text{Haar}(n)$, where we choose to work in an $n$-qubit space. We can expand any unitary as a linear combination of Pauli strings:
\begin{equation}
U = \sum_{i \in [4^n]} c_iP_i,
\end{equation}
where $P_i \in \{\bigotimes\limits_{j\in[n]}\sigma_j\}$ and $\sigma_j \in \{\sigma_x, \sigma_y, \sigma_z, \sigma_0\}$. A core attribute of the Haar distribution is that in expectation, there will be no bias toward any particular Pauli string. As a result, 
\begin{equation}
    \mathop{\mathbb{E}}\limits_{U}[|c_i|^2]=\frac{1}{4^n},
\end{equation}
since there are $4^n$ $n$-qubit Pauli strings. Additionally, it is clear that $c_i \propto \Tr[P_iU]$, which implies that $|c_i|^2$ can be expanded by a $(1,1)$ polynomial. As such, this property is preserved for any unitary $k$-design for $k\geq1$, not just a Haar-random unitary  \cite{twodesigntwirl2009}.

\begin{figure}[]
    \centering
    \includegraphics[width=0.9\linewidth]{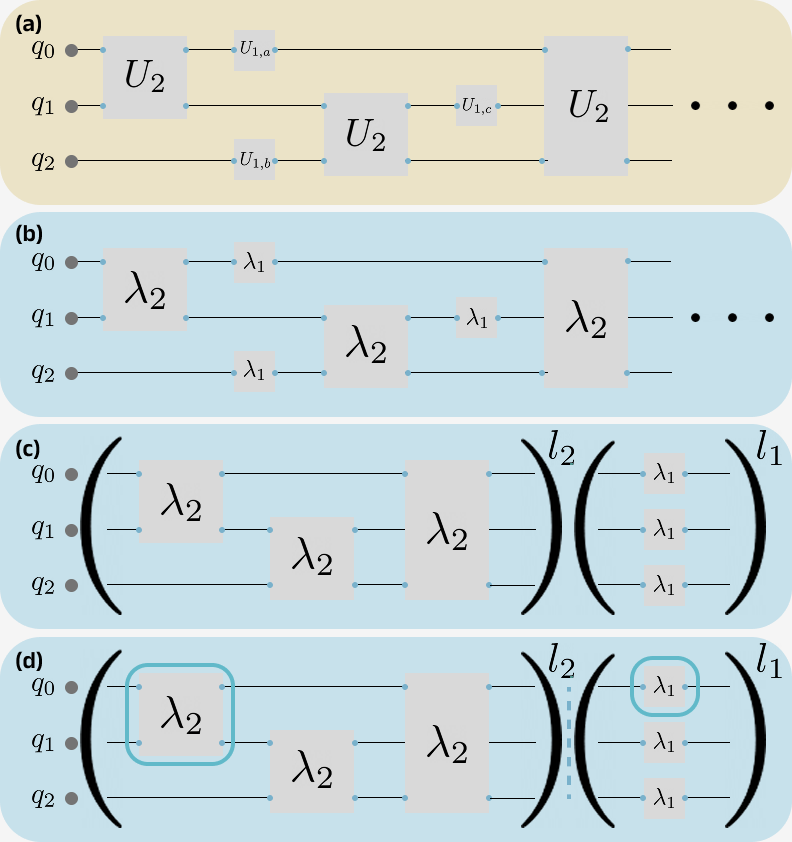}
    \caption{Our approach for deriving each noisy gate's contribution to global depolarization. (a) Assume we have some circuit, each of whose qubit pairs experiences the same number of two-qubit gates, and each of whose qubits experiences the same number of single-qubit gates. (b) We can write the local depolarizing elements under our assumption that single-qubit gates all have depolarizing parameter $\lambda_1$ and two-qubit gates have depolarizing parameter $\lambda_2$. (c) Depolarizing error channels commute, so we can gather the two-qubit depolarizing elements in blocks of fully-connected motifs; likewise, we gather single-qubit depolarizing elements in complete layers of gates. In this case, we find that our full circuit is composed with $l_2$ two-qubit gates per qubit pair, and $l_1$ single-qubit gates per qubit.  (d) By analyzing the global depolarization due to a fully-connected motif of two-qubit gates, we can find the average contribution per gate. We will ignore the error incurred by single-qubit gates, as they often have much smaller error than two-qubit gates.}
    \label{fig:derivation_figure}
\end{figure}
\subsection{Two-Qubit Gate Error \label{sec:global_twoq_depolarization}}
The form of a global depolarizing element on an $n$-qubit register is:
\begin{equation} \label{eq:globaldepol}
\mathcal{E}_n(\rho)=\lambda\rho+(1-\lambda)\frac{I}{2^n},
\end{equation}
which attenuates off-diagonal elements of the density matrix. Local two-qubit depolarizing elements will have the form:
\begin{equation}
\mathcal{E}_2^{(i,j)}(\rho)=\lambda_2 \rho + (1-\lambda_2)\frac{I_{i,j}}{4} \otimes \mathrm{Tr}_{i,j}(\rho),
\end{equation}
behaving the same way, except that they act on the two-qubit density submatrix labelled by $i,j$. For now, we assume that all two-qubit elements have the same depolarizing parameter and that all qubit pairs have the same number of two-qubit gates; we will discuss how to relax these assumptions in Section \ref{sec:circuit_balancing}. Under these assumptions, we can proceed with the analysis depicted in Figure \ref{fig:derivation_figure}. In particular, it will suffice to analyze a single maximally connected two-qubit gate motif and then find the average contribution per gate. 

We note that an implicit assumption made in this analysis is the commutativity of the local depolarizing elements with surrounding circuit gates. Although this is not true in the general setting that may be driven by low-entanglement dynamics, it has been shown that scrambling ensembles such as random circuits contain sufficient high-entanglement dynamics to approximate local depolarizations as global operators, and the probability of subsequent error cancellation is minimized \cite{quantumsupremacy2019} \cite{supremacy2018}. Prior work has also argued and shown empirically that sufficiently scrambling circuits exhibit the property of local error channels acting globally \cite{prethermalization2023}. We will thus proceed under the assumption that the local depolarizing elements approximately commute with other circuit elements as well as with each other in this setting. 

Now we define the undirected connectivity graph associated with the target device topology as
\begin{equation}
    G=(V,E),
\end{equation}
where $V$ represents the set of vertices (qubits) and $E$ represents the set of vertex 2-tuples whose elements are connected on the device, \textit{i.e.}, $\{(v_1, v_2)\}$. The global error channel of the previously mentioned maximally connected motif over this connectivity graph can then be constructed as the composition of all local error channels:
\begin{equation}
\mathcal{E}_{c,2}(\rho) = \mathop{\bigcirc}_{(i,j)\in E}\mathcal{E}^{(i,j)}_2(\rho).
\end{equation}
One of the defining characteristics of the global depolarizing error channel, $\mathcal{E}_n$, is that it attenuates non-identity Pauli strings uniformly, agnostic to any feature of the string itself. However, applying $\mathcal{E}_{c,2}$ to a Pauli string $P_i$ results in a string-dependent attenuation. If we define $\lambda_{c,2}$ as the global attenuating factor for the fully connected motif of two-qubit depolarizing errors and $N_2(P_i)$ as the number of device-connected qubit pairs on which $P_i$ acts nontrivially, then it is clear that
\begin{equation}\label{eqn:motifdepolarization}
    \lambda_{c,2}(P_i)=\lambda_2^{N_2(P_i)}.
\end{equation}
As discussed in Section \ref{sec:pauli_support}, the Pauli support strings are uniformly distributed over unitary $k$-designs. As a result, it will be instructive to model $N_2$ as a random variable over uniformly distributed Pauli strings. We will now derive the expectation and variance of $N_2$ over equiprobable Pauli strings. 

In order to do this, we will construct indicator random variables associated with each device-connected qubit pair:
\begin{equation}
    X_{(v_1,v_2)}(P_i)=\begin{cases}
0 \text{ if $P_i$ acts trivially on $(v_{1},v_{2})$},\\
1 \text{ else}.\\
\end{cases}
\end{equation}
Then clearly, we can write $N_2(P_i)$ as a sum over the indicator variables associated with the device-connected qubit pairs:
\begin{equation}
    N_2(P_i)=\sum_{(v_1,v_2)\in E}X_{(v_1,v_2)}(P_i).
\end{equation}
We will denote the uniform distribution of possible Pauli strings as $U_P$. In order to find the expectation of $N_2$, we start with the expectation over each indicator variable. The probability of a nontrivial interaction is complementary to the probability of both indices $v_1$ and $v_2$ being $I$. Sampling each constituent Pauli individually replicates the uniform distribution over all Pauli strings, so the probability of each index being $I$ is exactly $\frac{1}{4}$. As a result, for the two-qubit edge variable $X_{(v_1,v_2)}(P_i)$, we have
\begin{equation}
     \mathop{\mathbb{E}}_{P_i \sim U_P}[X_{(v_1,v_2)}(P_i)]=\frac{15}{16}.
\end{equation}
Since the indicator expectation is constant, then by linearity, we can determine the expectation of $N_2$ as
\begin{equation}\label{eqn:expectationresult}
    \mathop{\mathbb{E}}_{P_i \sim U_P}[N_2(P_i)]=\frac{15}{16}|E|,
\end{equation}
where $|E|$ denotes the size of the set $E$.
Now, in order to analyze the stability of $N_2(P_i)$, we calculate its variance under $P_i\sim U_P$. By definition, we have
\begin{equation}\label{eqn:varexpansion}
\begin{split}
    \mathop{\text{Var}}_{P_i\sim U_P}[N_2(P_i)]=\sum_{(v_1,v_2)\in E}\mathop{\text{Var}}_{P_i\sim U_P}[X_{(v_1,v_2)}]\\+\sum_{\substack{(v_1,v_2)\in E\\(v_1',v_2')\in E\\(v_1,v_2)\neq (v_1',v_2')}}\mathop{\text{Cov}}_{P_i\sim U_P}[X_{(v_1,v_2)},X_{(v_1',v_2')}].
    \end{split}
\end{equation}
The first summation is straightforward. For any particular indicator variable, we have already shown that over $P_i\sim U_P$, its probability is $\frac{15}{16}$ to take on value $X_{(v_1,v_2)}=1$ and $\frac{1}{16}$ to take on value $X_{(v_1,v_2)}=0$. As a result, we have
\begin{equation}\label{eqn:indicatorvariance}
\begin{split}
    \mathop{\text{Var}}_{P_i\sim U_P}[X_{(v_1,v_2)}]=\Pr_{P_i\sim U_P}[X_{(v_1,v_2)}(P_i)=0] \\\cdot \Pr_{P_i\sim U_P}[X_{(v_1,v_2)}(P_i)=1]\\=\frac{15}{256}.
    \end{split}
\end{equation}
We now turn our attention to the covariance summation. We can break this into two cases: $(v_1,v_2)$ and $(v_1',v_2')$ are adjacent (meaning that the two edges have a qubit in common), or they are disjoint. In the latter case, the random variables $X_{(v_1,v_2)}$ and $X_{(v_1',v_2')}$ are independent and thus have zero covariance. Thus, we need only consider the covariance summation over pairs of adjacent edges in our connectivity graph. In this case, the indicator variables are no longer independent. Without loss of generality, we will assume that $v_1$ is the shared qubit, such that
\begin{equation}\label{eqn:covariance}
\begin{split}
    \mathop{\text{Cov}}_{P_i\sim U_P}[X_{(v_1,v_2)},X_{(v_1,v_2')}] = \mathop{\mathbb{E}}_{P_i \sim U_P}[X_{(v_1,v_2)}X_{(v_1,v_2')}]\\-\mathop{\mathbb{E}}_{P_i \sim U_P}[X_{(v_1,v_2)}]\mathop{\mathbb{E}}_{P_i \sim U_P}[X_{(v_1,v_2')}].
    \end{split}
\end{equation}
Now, we know that
\begin{equation}
\begin{split}
    \mathop{\mathbb{E}}_{P_i \sim U_P}[X_{(v_1,v_2)}X_{(v_1,v_2')}]\\=\Pr_{P_i \sim U_P}[X_{(v_1,v_2)}=1 \cap X_{(v_1,v_2')}=1].
\end{split}
\end{equation}
Using the principle of inclusion-exclusion, we can rewrite this as
\begin{equation}\label{eqn:correlatedexpectation}
    \begin{split}
    \mathop{\mathbb{E}}_{P_i \sim U_P}[X_{(v_1,v_2)}X_{(v_1,v_2')}]\\=1-(\Pr_{P_i \sim U_P}[X_{(v_1,v_2)}=0] + \Pr_{P_i \sim U_P}[X_{(v_1,v_2')}=0]\\-\Pr_{P_i \sim U_P}[X_{(v_1,v_2)}=0 \cap X_{(v_1,v_2')}=0]).
\end{split}
\end{equation}
It is clear that $\Pr_{P_i \sim U_P}[X_{(v_1,v_2)}=0]=\Pr_{P_i \sim U_P}[X_{(v_1,v_2')}=0]=\frac{1}{16}$ from previous analysis. The last term, $\Pr_{P_i \sim U_P}[X_{(v_1,v_2)}=0 \cap X_{(v_1,v_2')}=0]$, represents the probability that two adjacent qubit pairs are acted on trivially by a Pauli string. Since there are only three independent qubits ($v_1$ is shared between the pairs), this is simply the probability that the Pauli string contains $I$ at indices $v_1$, $v_2$, and $v_2'$. Over $U_P$, this is simply $\frac{1}{4^3}$. Substituting this result into Equation \ref{eqn:correlatedexpectation}, we have
\begin{equation}\label{eqn:correlatedexpectationoutcome}
    \mathop{\mathbb{E}}_{P_i \sim U_P}[X_{(v_1,v_2)}X_{(v_1,v_2')}]=1-\frac{1}{16}-\frac{1}{16}+\frac{1}{4^3}=\frac{57}{64}.
\end{equation}
We can then substitute this result into Equation \ref{eqn:covariance} together with our existing results for the independent indicator expectations to produce
\begin{equation}\label{eqn:covresult}
     \mathop{\text{Cov}}_{P_i\sim U_P}[X_{(v_1,v_2)},X_{(v_1,v_2')}] = \frac{57}{64}-\frac{15}{16}\cdot\frac{15}{16}=\frac{3}{256}.
\end{equation}
We will now define the set of pairs of adjacent edges in our connectivity graph:
\begin{equation}
\begin{split}
    S = \{((v_1,v_2),(v_3,v_4))\mid ((v_1,v_2)\in E) \\\land  ((v_3,v_4)\in E) \\\land (v_1 \in \{v_3,v_4\} \veebar v_2 \in \{v_3, v_4\})\}.
\end{split}
\end{equation}
Combining this definition with the results from Equations \ref{eqn:covariance} and \ref{eqn:indicatorvariance} and substituting into Equation \ref{eqn:varexpansion} yields
\begin{equation}\label{eqn:varianceresult}
    \mathop{\text{Var}}_{P_i\sim U_P}[N_2(P_i)]=\frac{15}{256}|E|+\frac{3}{256}|S|,
\end{equation}
where $|S|$ denotes the size of the set $S$.
Now we have rigorously determined the mean and variance of $N_2$ as a random variable over $P_i\sim U_P$. As a result, via the Chebyshev inequality, we can state with arbitrarily high confidence that $N_2=\mathop{\mathbb{E}}_{P_i \sim U_P}[N_2(P_i)] \pm O(\sqrt{\mathop{\text{Var}}_{P_i\sim U_P}[N_2(P_i)]})$. 

We now return to Equation \ref{eqn:motifdepolarization}, where we wrote the global depolarization parameter for a maximally dense motif of gates. We can treat $N_2$ as a random variable over the uniform Pauli string distribution so that with arbitrarily high probability
\begin{equation}
    \lambda_{c,2} \approx \lambda_2^{\mathop{\mathbb{E}}_{P_i \sim U_P}[N_2(P_i)] \pm O(\sqrt{\mathop{\text{Var}}_{P_i\sim U_P}[N_2(P_i)]})}.
\end{equation}
This represents the depolarizing parameter corresponding to the entire maximally connected motif of two-qubit gates. In order to produce the average per-gate contribution within this motif, we simply take the $|E|^{\text{th}}$ root of $\lambda_{c,2}$:
\begin{equation}
    \lambda_{\text{avg},2Q} \approx \lambda_2^{(\mathop{\mathbb{E}}_{P_i \sim U_P}[N_2(P_i)] \pm O(\sqrt{\mathop{\text{Var}}_{P_i\sim U_P}[N_2(P_i)]}))/|E|}.
\end{equation}
We can now substitute the expectation and variance for $N_2$ derived in Equations \ref{eqn:expectationresult} and \ref{eqn:varianceresult} to produce
\begin{equation}
     \lambda_{\text{avg},2Q} \approx \lambda_2^{\frac{15}{16} \pm O(\sqrt{\frac{15}{16}|E|+\frac{3}{256}|S|})/|E|}.
\end{equation}
In order to understand the behavior of the variance asymptotics in the above estimate, we consider our connectivity graph, $G$, to be $d$-regular. As a result, $|E|\in O(nd)$. Further, each qubit mediates adjacency over $d \choose 2$ pairs of edges, so $|S| \in O(nd^2)$. Thus, for $d$-regular connectivity graphs, we find
\begin{equation}
    \lambda_{\text{avg},2Q} \approx \lambda_2^{\frac{15}{16} \pm O(\frac{1}{\sqrt{n}})},
\end{equation}
so that
\begin{equation}
    \lambda_{\text{avg},2Q} \underset{n >>1}{\approx} \lambda_2^{\frac{15}{16}}.
\end{equation}
As a result, in the large register limit ($n>>1$), any $d$-regular connectivity graph $G$ admits an average global depolarization per gate of $\lambda_2^{\frac{15}{16}}$ within a maximally connected lattice. 

\begin{figure}
    \centering
    \includegraphics[width=0.95\linewidth]{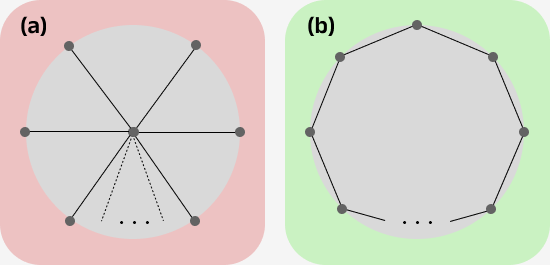}
    \caption{Example device connectivity graphs, where qubits are denoted by nodes, and connections by edges. In (a), qubits are connected in a star configuration; more qubits are added by radially connecting them to a central qubit. In (b), qubits are connected in a circular configuration.}
    \label{fig:connectivitygraphs}
\end{figure}

Additionally, this result holds for any quantum computer whose connectivity graph admits asymptotics such that $\sqrt{|S|}/|E| \rightarrow 0$ for large $n$. As such, we anticipate these results to hold for many common device topologies, such as linear qubit chains, heavy-hex lattices, and all-to-all connected devices. However, this is not the case for all possible connectivity graphs. For example, in Figure \ref{fig:connectivitygraphs}a, qubits are configured in a star connectivity graph, such that $|S|$ grows quadratically in $n$. In contrast, the two-regular graph in \ref{fig:connectivitygraphs}b is such that $|S|$ grows linearly in $n$. Intuitively, the depolarization that the central qubit experiences in the star connectivity is much greater than that experienced by the qubits on the circle, thus making the depolarization Pauli-dependent. However, the uniformity of depolarization in the two-regular connectivity graph produces a Pauli-independent depolarization in the large register limit.

The physical meaning of the value $\lambda_{\text{avg},2Q}$ is the average global depolarization per gate within a single dense motif. In order for this average depolarization metric to be useful, it is necessary to assume that the circuit is ``balanced.'' In other words, it must be the case that every depolarizing gate in the target circuit has the same depolarizing parameter, and that each qubit pair participates in roughly the same number of two-qubit gates. As previously stated, we will relax these assumptions in Section \ref{sec:circuit_balancing}.

\subsection{Circuit Balancing \label{sec:circuit_balancing}}
We have shown how it is possible to approximate the register-wide depolarization from a single noisy two-qubit gate based on its local depolarization under strong assumptions on the structure of the input circuit. We now discuss a method to circumvent these limitations, which we dub ``circuit balancing.'' Intuitively, a sequence of $s$ two-qubit gates with depolarizing parameter $\lambda$ can be considered a single depolarizing element with parameter $\lambda^s$. As such, if we have a target device with varying levels of two-qubit depolarization, then we simply choose to perform the fewest two-qubit operations with the noisiest gates. 

\begin{figure}[]
    \centering
    \includegraphics[width=0.9\linewidth]{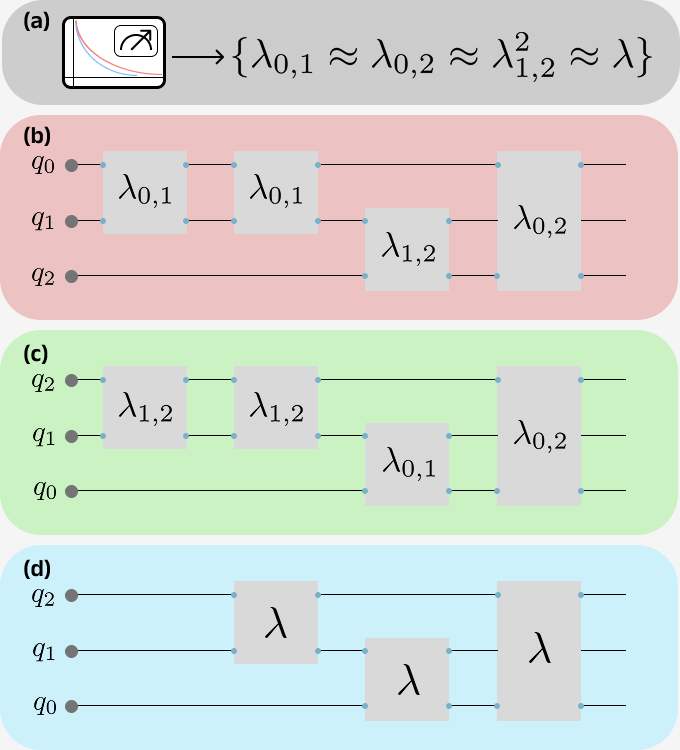}
    \caption{An example of circuit balancing. (a) Suppose we have conducted a routine set of benchmarking on a three-qubit device and found this approximate relationship over the depolarizing parameters corresponding to the pairwise two-qubit gates. (b) We are given a circuit using four two-qubit gates, which amount two four local depolarizations. Without circuit balancing, the total local depolarization on each two-qubit pair is different. (c) By permuting the qubit labels in the register (which is simply a form of layout selection for an all-to-all connected device), we enforce similar total local depolarization across qubit pairs. (d) Using the relationships determined through benchmarking in (a), we can combine adjacent local depolarizations to see that we have now satisfied our assumptions of each qubit pair being involved in the same number of depolarizing elements with the same parameter.}
    \label{fig:circuit_balancing}
\end{figure}

In other words, we want the total local depolarization on each qubit pair to be roughly the same, which we can control based on layout selection. Assume that we have $n_p$ physical qubits and our circuit contains $n$ logical qubits. Then, a layout selection can be defined as an injective mapping between logical, circuit-level qubits $[n]$ and physical qubits $[n_p]$; in particular, as $F:[n] \rightarrow [n_p]$. We will define the set of valid layouts $L$ for a circuit $c$ as $L(c)$, such that for all $F \in L$, $F(i)$ and $F(j)$ are connected on the target quantum computer when there is a two-qubit gate on qubits $(i,j)$ in our logical circuit. We can define the number of two-qubit gates in our logical circuit between logical qubits $i$ and $j$ as $G(i,j)$, each of which will have a physical depolarization under layout selection $F$, given by $\lambda_{(F(i),F(j))}$. Then, in order to find the total local two-qubit depolarization over pair $(i,j)$, we simply raise the depolarizing parameter of the physical pair to the power of the number of two-qubit gates on that pair:
\begin{equation}
    \lambda_{c,\text{local},(i,j)}=\lambda_{(F(i),F(j))}^{G(i,j)}.
\end{equation}
In order for our assumptions to hold, we simply want $\lambda_{c,\text{local},(i,j)}$ to be as constant as possible over all pairs $(i,j)$. We can formulate this consideration as an optimization problem over possible layouts. As a result, we simply have
\begin{equation}\label{eqn:optimization}
    F^* = \underset{F\in L}{\arg\min} [\underset{(i,j)\in E }{\text{Var}}[\lambda_{(F(i),F(j))}^{G(i,j)}]].
\end{equation}
The mapping $F^*$ represents the layout selection that produces the smallest variance in total local depolarization induced by two-qubit gates. We note that we can approximate solutions to this problem using techniques like simulated annealing even for relatively large register sizes, and this objective may easily be incorporated as a consideration into existing heuristic-based layout selectors \cite{sabre2019}. The search space may grow up to combinatorially (for all-to-all connectivity) over increasing register size, but even approximate solutions with a fixed iteration count can yield positive results for moderately sized registers, as shown numerically in Section \ref{sec:results}. Additionally, we demonstrate that evaluating the circuit balancing cost function over a small set of preselected layouts can lead to more physically meaningful estimates of depolarizing error on real hardware in Section \ref{sec:hardware_results}. For a toy example of solving a three-qubit circuit balancing layout selection, where we optimize over the layout $F$, see Figure \ref{fig:circuit_balancing}.

After this optimization problem is approximately solved, we can define an arbitrary $\lambda_{\text{target}}$ (for example, $0.999$), and find the number of equivalent gates per pair with this depolarizing parameter by solving for $k_{(i,j)}$: $\lambda_{\text{target}}^{k_{(i,j)}} = \lambda_{(i,j)}^{G(i,j)}$. Now, since each qubit pair undergoes roughly the same depolarization, we can apply the previous analysis using $\lambda_{\text{target}}$ as the same depolarizing parameter for each qubit pair and $k_{(i,j)}$ as the number of gates on each pair.

\subsection{Global Depolarization Estimation Procedure \label{sec:global_parameter_estimation}}
We will now assume that two-qubit gate error dominates global depolarization, neglecting the contributions of single-qubit gate error. It is often the case that single-qubit gate depolarization is several orders of magnitude lower than two-qubit gate depolarization \cite{leblond2025}. Future work may build on our approach by heuristically including the placement of single-qubit gates in the circuit balancing cost function. 

Under this assumption, we summarize our estimation procedure:
\begin{enumerate}
    \item Given a circuit implementing a unitary $k$-design and a quantum computer with benchmarked two-qubit gate depolarizing error, approximate a solution to the optimization problem  presented in Section~\ref{sec:circuit_balancing} above. Using a fixed $\lambda_{\text{target}}$, find the number of equivalent gates per qubit pair, $k_{(i,j)}$.
    \item Since the circuit has been balanced, then each qubit pair experiences roughly the same number of equivalent gates with parameter $\lambda_{\text{target}}$. As such, each gate contributes a global depolarization parameter of approximately $\lambda_{\text{target}}^{\frac{15}{16}}$. 
    \item Now, we can estimate the register-wide depolarizing parameter, $\lambda_n$, by counting the contribution from every equivalent two-qubit gate: $\lambda_n \approx \lambda_{\text{target}}^{\frac{15}{16}\sum\limits_{(i,j)\in E}k_{(i,j)}}$.
\end{enumerate}
\section{Depolarizing Noise Inversion\label{sec:inversion}}
\subsection{Problem Setting and Procedure \label{sec:procedure}}
In an experimental setting, we can assume without loss of generality (assuming no mid-circuit measurements) that we run a circuit on a target quantum computer, measure the entire register in the Z-basis, and produce a Z-basis bitstring distribution marginalized over the shot count as the output. We will denote the output distribution as $\mathcal{D} = \{b_i,p_i\}$, where $b_i$ represents a particular Z-basis bitstring, and $p_i$ represents the probability of observing it. Operating within this bitstring picture, if we have a target bitstring $b_t$ whose noiseless probability of observation is $p_t$, then with a global depolarizing parameter $\lambda$, the empirical probability will be given by the survival function on $p_t$:
\begin{equation}
    S(\lambda,p_t)=(p_t-\frac{1}{2^n})\lambda + \frac{1}{2^n},
\end{equation}
which follows from the definition of depolarizing noise on a Z-basis bitstring observable (see Equation~\ref{eq:globaldepol}). Intuitively, depolarizing noise attempts to maximally mix a quantum state, so for small $\lambda$ values, the measured bitstrings will appear to be increasingly equiprobable. If we have $m$ global depolarizing elements, then this process decays bitstring probabilities exponentially:
\begin{equation}
    S_m(\lambda,p_t)=(p_t-\frac{1}{2^n})\lambda^m + \frac{1}{2^n}
\end{equation}
Notably, in the large register limit ($n >> 1$), $S(p_t)$ is linear and can be written as 

\begin{equation}
    S_m(\lambda,p_t) \approx \lambda^m p_t.
\end{equation}
Now, in order to be in the regime where we can simply invert $S(p_t)$ on each element of $\mathcal{D}$, we need to sufficiently suppress coherent error because this can produce significant perturbations around $S(p_t)$, hindering faithful reconstruction of the noiseless distribution. 
We note that Pauli twirling of a two-qubit gate can be done without incurring any two-qubit gate overhead. 
Our procedure for a quantum circuit to be executed on noisy quantum hardware is then as follows:
\begin{enumerate}
    \item Neglecting single-qubit gate error, calculate the effective global depolarizing parameter, $\lambda_n$, for the target circuit via the procedure given in Section \ref{sec:global_parameter_estimation} after circuit balancing as in Section~\ref{sec:circuit_balancing}.
    \item Produce $t$ Pauli twirls (each over all two-qubit gates) for the target circuit and run them on the target quantum computer. Produce an average distribution $\mathcal{D}_{\text{noisy, avg}} = \{b_i, p_i\}$.
    \item Invert the depolarizing noise to produce an estimate of the noiseless distribution $\mathcal{D}' = \{b_i,S^{-1}(\lambda_n,p_i)\}$ using the estimated $\lambda_n$.
    \item If there is a slight error in our $\lambda_n$ estimation process, then $\mathcal{D}'$ may not be a valid probability distribution. As such, the final step is to produce $\mathcal{D}^*=\text{proj}_C \mathcal{D'}$, where $C$ is the probability simplex of appropriate dimension. We anticipate (and show empirically in Section \ref{sec:numerical_experiments}) that the circuit balancing objective is a good proxy for $\lambda_n$ estimation error, so appropriate layout selection mitigates against large estimation error.
\end{enumerate}
Note that Pauli twirling of two-qubit gates incurs only single-qubit-gate and shot overhead; no two-qubit gate overhead is incurred. We will now derive an expression to evaluate the asymptotic scaling of the required number of twirls, $t$, in terms of our error tolerance, device coherent error, and depolarizing parameter.

\subsection{Twirl Overhead Estimate \label{sec:twirl_overhead}}
In order to determine the number of twirls needed, we begin by defining measurement operators $M_i$ corresponding to observation of Z-basis bitstrings $b_i$, with eigenvalues 0 and 1. Additionally, we define the quantum error channel with both coherent and depolarizing noise $\mathcal{E}_{C+D}$ and that with just depolarizing noise $\mathcal{E}_D$. Then, for a particular Pauli twirl $T_j$, we can define a random variable representing the probability of observing bitstring $b_i$:
\begin{equation}
    Y_j = \Tr[M_i \mathcal{E}_{C+D} ^{(T_j)}(\rho)].
\end{equation}
Pauli twirling produces an unbiased estimate of a target expectation in the absence of coherent error, so we have
\begin{equation}
    \mathop{\mathbb{E}}[Y_j]=\Tr[M_i \mathcal{E}_{D}(\rho)].
\end{equation}
Going forward, we will use $p_{i,\text{DO}}$ to denote the probability of measuring bitstring $b_i$ with only depolarizing error such that
\begin{equation}
    p_{i,\text{DO}} = \Tr[M_i \mathcal{E}_{D}(\rho)].
\end{equation}
However, each such $Y_j$ is drawn from a distribution with nonzero variance. The setting of Pauli twirling noisy gates is analogous to extracting infidelity via randomized benchmarking. It has been shown in randomized benchmarking literature that in a circuit with $m$ noisy elements, the variance of expectation estimates over randomized compiling will be $O(mr)$, where $r$ is the infidelity of each noisy element \cite{wallman2014}. We will assume small coherent rotations as the source of coherent error. For small coherent rotations of angle $\theta$, the infidelity scales with $\theta^2$, so the expectation variance over Pauli twirls will be $O(m \theta^2)$. 

By averaging over $t$ twirls, we can denote the twirled estimate as
\begin{equation}
    \hat{p}_{i,\text{DO}}=\frac{1}{t}\sum_{j=1}^t Y_j.
\end{equation}
We use the notation $\hat{p}_{i,\text{DO}}$ since Pauli twirling produces an unbiased estimate of the error-free probability, so that averaging over $Y_j$ produces an estimate of the depolarizing-only probability. Applying the Central Limit Theorem, since the variance of $Y_j$ is $O(m\theta^2)$, we have
\begin{equation}
    \hat{p}_{i,\text{DO}}\sim \mathcal{N}({p}_{i,\text{DO}},\frac{O(m\theta^2)}{t}).
\end{equation}
We can then define a random variable that denotes the difference between expectation of our twirled estimate in the presence of both depolarizing and coherent error and the expectation under only depolarizing error:
\begin{equation}
    \Delta_{\text{PR}} = \hat{p}_{i,\text{DO}}-{p}_{i,\text{DO}}.
\end{equation}
Since the second term is constant with respect to twirls, we can subtract it from the mean of $\hat{p}_{i,\text{DO}}$ to write $\Delta_{\text{PR}}$ as an independent random variable:
\begin{equation}
    \Delta_{\text{PR}} \sim \mathcal{N}(0,\frac{O(m\theta^2)}{t}).
\end{equation}
Our goal is to determine the quantity
\begin{equation}
    \delta = |S_m^{-1}(\lambda,\hat{p}_{i,\text{DO}}) - {p}_{i,\text{ideal}}|,
\end{equation}
which tells us the difference in bitstring probabilities between the ideal and noisy settings, when we invert depolarizing noise. Assuming perfect knowledge of the depolarizing parameter $\lambda$, we can rewrite this as
\begin{equation}
    \delta = |S_m^{-1}(\lambda, \hat{p}_{i,\text{DO}}) - S_m^{-1}(\lambda,{p}_{i,\text{DO}})|.
\end{equation}
We can apply the large register limit to approximate $S$ as linear, so that this becomes
\begin{equation}
    \delta = \frac{1}{\lambda^m}|\hat{p}_{i,\text{DO}}-{p}_{i,\text{DO}}|,
\end{equation}
which is just
\begin{equation}
    \delta = \frac{1}{\lambda^m}|\Delta_{\text{PR}}|.
\end{equation}
We can then use the properties of the normal distribution to obtain
\begin{equation}
    \delta \sim |\mathcal{N}(0,\frac{O(m\theta^2)}{t \lambda^{2m}})|.
\end{equation}
Using standard Gaussian tail bounds, we have
\begin{equation}
    \delta^2 \leq z_c^2 \frac{O(m\theta^2)}{t \lambda^{2m}},
\end{equation}
for some z-score $z_c$ associated with a confidence of $1-2e^{-z_c^2/2}$. Absorbing the confidence into our asymptotic analysis and solving for $t$ then yields the following bound on the required number of twirls $t$ for the error mitigation protocol:
\begin{equation}
    t \geq O(\frac{m\theta^2}{\delta^2 \lambda^{2m}}).
\end{equation}

In order to understand this bound within the context of circuit balancing, we can interpret the parameter $\lambda$ in the above inequality as corresponding to $\lambda_{\text{target}}^{\frac{15}{16}}$, in which case $\theta$ corresponds to the worst case coherent error from the largest block of physical gates that corresponds to a single instance of our depolarizing element with local parameter $\lambda_{\text{target}}$. For example, if the qubit pair with the lowest depolarizing error has local parameter $\lambda_{\text{best}}$ per physical gate, then the number of effective target gates will be given by $g_{\text{best}}$ in $\lambda_{\text{best}}^{g_{\text{best}}} =\lambda_{\text{target}}$. $\theta$ will then correspond to the coherent error of the block of $g_{\text{best}}$ gates (assuming the coherent error per gate is similar). Finally, the number of noisy gates $m$ will be replaced by the number of effective gates, $\sum\limits_{(i,j)\in E}k_{(i,j)}$. Note that if the two qubit depolarizing parameters of the device are all very similar, say close to $\lambda_d$, and the circuit is naturally balanced (each qubit undergoes a similar number of two-qubit operations), then we can choose $\lambda_{\text{target}}=\lambda_d$. Then, $g_{\text{best}} \rightarrow 1$ and $\sum\limits_{i<j}k_{(i,j)} \rightarrow m$. In this setting we arrive at a bound in terms of the gate-level depolarization given by
\begin{equation}
    t \geq O(\theta^2{\delta^{-2}\lambda_{d}^{-{\frac{15}{16}}2m}}m).
\end{equation}

\section{Numerical Experiments\label{sec:numerical_experiments}}
\subsection{Methods}
\subsubsection{Depolarizing Noise Only: Scaling System Size\label{sec:depolarizing_only_methods}}

To study the sensitivity of the depolarizing noise correction method to system size, for each $7\leq n \leq 11$, we construct a simulated all-to-all connected quantum computer whose two-qubit gates have local, two-qubit depolarizing parameters sampled uniformly from $[0.999,1]$. We choose CNOT, X, SX, and RZ gates as the simulated device basis gate set. We enforce symmetric depolarizing errors such that the depolarizing parameter for CNOT(i,j) is equal to that of CNOT(j,i). We assume access to the depolarizing parameters for individual two-qubit gates via benchmarking. Since we want to test only the depolarizing noise diagnosis and mitigation in this test, we will not introduce any coherent error into this set of simulations.

For each such $n$, we produce $10$ approximate unitary 2-designs over the entire register, such that the number of logical qubits $n$ is equal to the number of physical qubits $n_p$. We choose to use approximate unitary 2-designs instead of constructing $n$-qubit unitaries directly because decomposing an $n$-qubit unitary into single- and two-qubit gates remains intractable even for small $n$. It is known that random quantum circuits $\epsilon$-approximate unitary $k$-design in linear depth \cite{rqcpoly2022}. 
Additionally, previous numerical studies have shown that random quantum circuits constructed over planar connectivity graphs can admit Porter-Thomas probability distributions, characteristic of Haar-random unitaries, in depth less than 30 for over 30 qubits \cite{supremacy2018}. As a result, we would expect even devices with relatively sparse connectivity to admit fast convergence to a unitary 2-design; since we simulate an all-to-all connected device, relatively shallow circuits will suffice for convergence to a unitary 2-design. We choose to use a random circuit sampling technique such that each layer of circuit construction consists of applying single-qubit Haar-random unitaries register-wide followed by an application of two-qubit Haar random unitaries to a maximally parallelized set of random qubit pairs. Given that our maximum simulation size is $n=11$, we choose a highly conservative fixed random circuit depth of $150$ for all of our simulations.

For each approximate 2-design, we transpile the sequence into the simulated basis gates using Qiskit level 3 (L3) transpilation \cite{qiskit}. We apply the Circuit Balancing technique discussed in Section \ref{sec:circuit_balancing} via simulated annealing, with $5000$ total iterations allowed for each $n$. By not scaling this iteration count with register size, we limit the classical overhead, but potentially at the expense of satisfying assumptions necessary to predict circuit-wide depolarization. 

Since there is no coherent error in this simulation, Pauli twirling is not necessary. For each of the $10$ circuits per register size $n$, we execute it on the simulated device and on an ideal simulator; we then compare the Z-basis probability distributions via the Hellinger distance, which is immune to divergence when there are zero probabilities present (Figure \ref{fig:infidelity_comparison}). The Hellinger distance from the ideal distribution (which we will subsequently refer to as simply ``Hellinger distance'' for brevity) is a proxy for infidelity and has been widely used in many quantum computing applications for this reason \cite{fidelitymetrics}. We do not simulate shot noise in this case because, for a fixed shot count, this will have an increasingly deleterious effect as $n$ increases and our goal here is to study the efficacy of depolarization estimation alone. We will also assume that the depolarization parameter lies in a regime such that attenuation of bitstring probability does not incur a significant shot overhead for necessary resolution.

\subsubsection{Depolarizing Noise Only: Scaling Error Asymmetry} \label{sec:asymmetry}
One drawback of our approximate unitary 2-design approach is that circuits will naturally have similar numbers of two-qubit gates on each qubit pair. As a result, there is very little additional optimization that circuit balancing can achieve. In order to stress test the resilience of our circuit balancing method to circuit asymmetry and device error variance, we construct a set of $10$ full random unitaries for $n=6$ qubits (the maximum size for which we can tractably perform this experiment). For each six-qubit unitary, we transpile it into the same basis gates as in Section \ref{sec:depolarizing_only_methods}. We expect there to be more variation in gates per qubit pair due to transpiler optimizations. 
We then construct $10$ simulated quantum computers by the same method as in Section \ref{sec:depolarizing_only_methods}, except that we now sample gate depolarizing parameters uniformly per simulation from $[1-\epsilon, 1]$, where $\epsilon$ (which we refer to as the depolarizing error bound), is varied in steps of $0.001$. 
We then find the mean Hellinger distance across the ten circuits for each simulated device in three settings: without mitigation, with mitigation but without circuit balancing, and with both mitigation and circuit balancing (Figure \ref{fig:variance_test}). 

\subsubsection{Depolarizing and Coherent Error}
Lastly, in order to test our method against coherent error, we take a similar approach to Section \ref{sec:depolarizing_only_methods}, except that we fix $n=10$ and reproduce experiments with varying amounts of coherent error. We introduce coherent error via a fixed XX-rotation after each CNOT gate, parametrized by angle $\theta$.
We choose XX-rotations as our coherent error source, due to miscalibrations in trapped-ion two-qubit gates often manifesting as unwanted XX-rotations \cite{leblond2025}. We use coherent error values $\theta \in \{3\cdot10^{-3},6\cdot 10^{-3},9\cdot 10^{-3}\}$ in units of radians.
We follow the same circuit generation process, now using $10$ circuits per coherent error strength, and once again analyze the Hellinger distance. We compare the unmitigated results with two sets of mitigated results: one set without twirling ($t=1$), and one set where we increase $t$ with the coherent error strength such that $t \in \{1,4,9\}$ (Figure \ref{fig:coherent_error}). Since the Hellinger distance depends strictly on bitstring probability error~\cite{fidelitymetrics}, we would expect that in order to maintain a roughly constant Hellinger distance, $t$ would need to increase quadratically in the coherent error rotation angle according to the bounds derived in Section \ref{sec:twirl_overhead}. We chose the number of twirls quadratically in $\theta$ for this reason. The proportionality constant is such that $1$ twirl corresponds to the coherent error rotation of $3\cdot 10^{-3}$ radians. However, rather than demonstrating an absolute bound on twirls for a desired error, we hope to empirically illustrate the stability of our protocol with the number of twirls chosen quadratically in $\theta$.

\subsection{Results \& Analysis \label{sec:results}}
\subsubsection{Depolarizing Noise Only: Scaling System Size \label{sec:scaling_size_results}}
\begin{figure}[]
    \centering
    \includegraphics[width=1.0\linewidth]{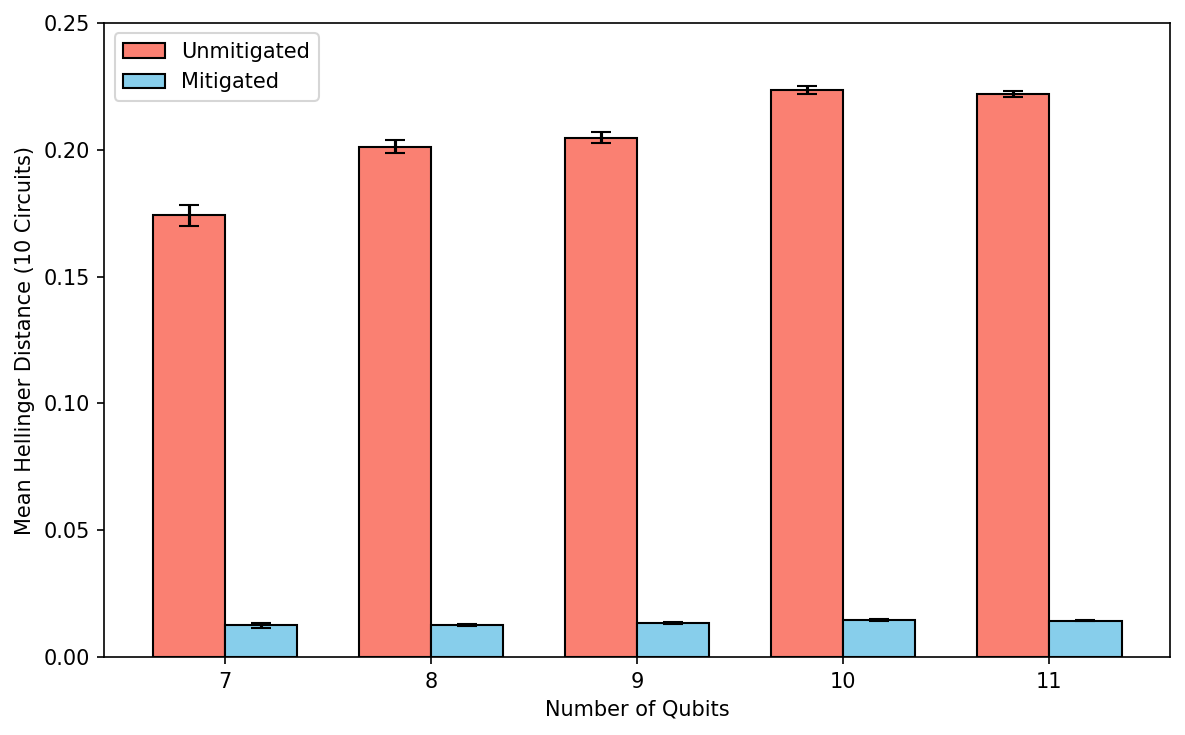}
    \caption{We construct $10$ approximate unitary 2-designs for several register sizes (depicted on the x axis), and we plot the mean Z-basis measurement distribution Hellinger distance from the ideal (depicted on the y axis; lower is better) for both the mitigated and unmitigated methods. Mean distance is reduced by at least $92\%$ in each case. The error model here contains only depolarization (no coherent errors).}
    \label{fig:infidelity_comparison}
\end{figure}

First, in the absence of coherent error, we present the mean Hellinger infidelities over various register sizes in Figure \ref{fig:infidelity_comparison}.

Although the depth for each register size is kept fixed, the number of two-qubit gates per layer increases linearly in the number of qubits, so we would expect the Hellinger distance without error mitigation to steadily increase. This is borne out by the increasing height of the pink columns. However, we see that the mitigation technique introduced here is able to nonetheless significantly suppress mean Hellinger distance (blue columns). Additionally, the mitigated Hellinger distance appears relatively stable over the register sizes tested, even though the unmitigated Hellinger distance does generally grow. We observe over $92\%$ mean Hellinger distance reduction for every register size tested. 
We also note that, as predicted above (Section~\ref{sec:asymmetry}), since the approximate 2-designs have very similar gate counts per qubit pair, there is little room for circuit balancing to improve the results.
Indeed, we see that on average, circuit balancing reduces Hellinger distance by less than $3\%$ on average in this setting. However, we expect to see different results on asymmetric circuits, and we probe this now.

\subsubsection{Depolarizing Noise Only: Scaling Error Asymmetry}
\begin{figure} \label{fig:variancetest}
    \centering
    \includegraphics[width=0.9\linewidth]{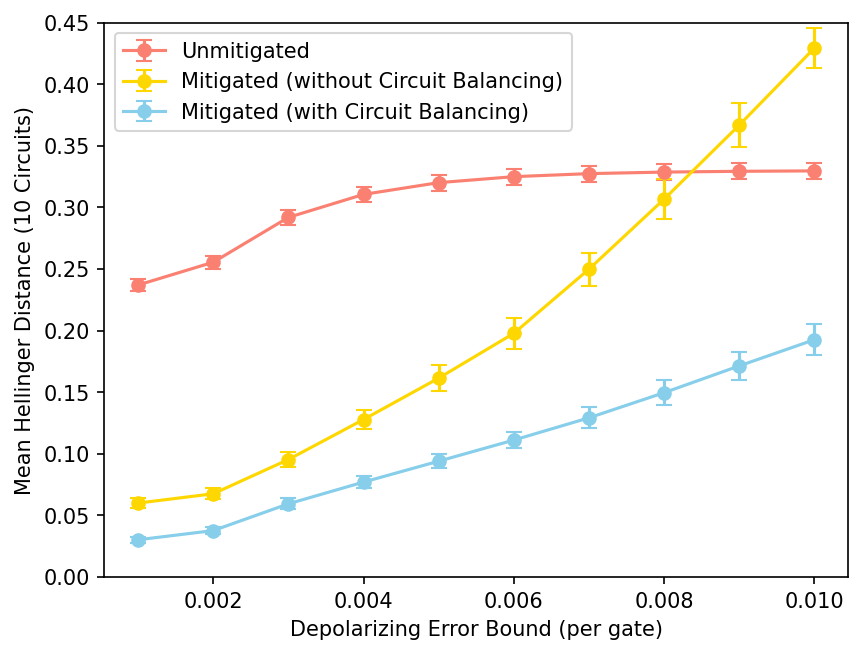}
    \includegraphics[width=0.9\linewidth]{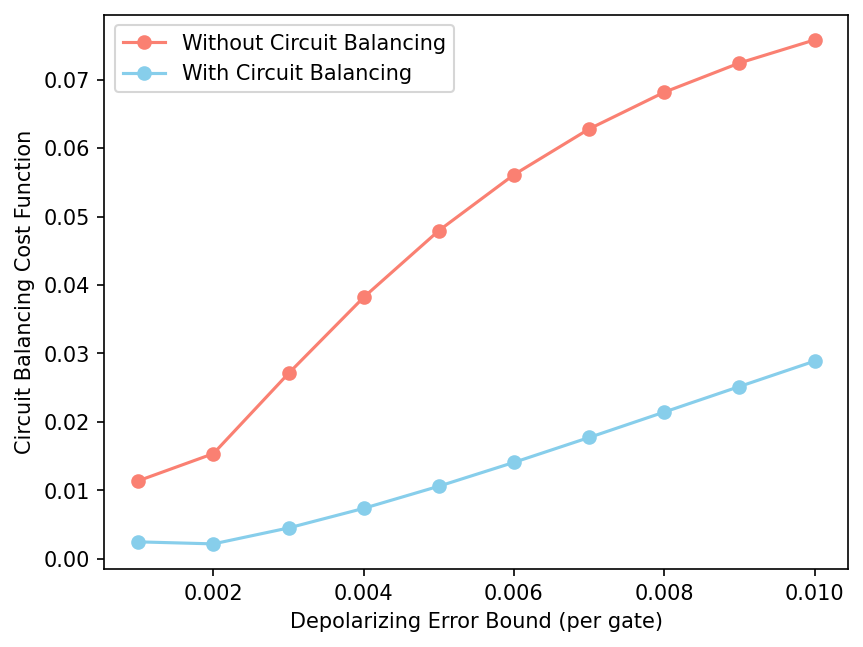}
    \caption{A comparison of methods over $10$ asymmetric Haar-random circuits, run on $10$ simulated backends with different error profiles. {In this case, the error models contain only gate-based depolarization (no coherent error), but the uniform distribution from which error per gate is selected broadens as the bound on the depolarizing error increases along the x axis.} Top: the mean Hellinger distance from the ideal (lower is better) of the three methods are compared over various gate error ranges for 6-qubit random unitaries. Even for fixed circuit balancing optimization, there is a broad error regime where mitigation with circuit balancing outperforms the default and is more resilient to error variance than without circuit balancing. Bottom: the circuit balancing cost function (Equation \ref{eqn:optimization}) with and without optimization. }
    \label{fig:variance_test}
\end{figure}

We now construct $10$ random six-qubit unitaries and use Qiskit L3 transpilation to produce circuits with large gate asymmetries. The transpiler produced the same gate counts for each unitary, indicating that it may have found a universal representation of a six-qubit unitary represented in a fixed circuit structure. 
This circuit structure exhibits the large asymmetry we expected, with a standard deviation of $>148$ CNOT gates over the qubit pairs. 
Over the $10$ tested simulators each tested with a $10$-circuit ensemble, we obtain the results summarized in Figure \ref{fig:variance_test}.

The top panel in Figure \ref{fig:variance_test} shows that as we increase the upper bound of the uniform distribution from which depolarizing error per gate is drawn, the Hellinger distance for all three methods grows. For the unmitigated runs (red symbols and connecting line), this effect is due to increased expectations of average gate error.
However, for the two mitigated methods, the increasing distance is due to errors in estimating the global depolarizing parameter. 
This is because when we increase the upper bound on the uniform error distribution, we increase both its expectation and variance. 
With large error variance, we expect an arbitrary layout selector to be unlikely to accidentally produce a balanced circuit.
Additionally, with large error expectation, we expect changes in qubit ordering to have greater effects on performance. 
We see empirically that in this situation, mitigation with circuit balancing (blue symbols and connecting line) provides a significant advantage over unmitigated runs over a large error regime and is more resistant to gate error spreading than mitigation without circuit balancing (yellow symbols and connecting line). 
Additionally, it is noteworthy that there is a regime in which mitigation without circuit balancing performs more poorly than the unmitigated approach.
This indicates that the error variance has sufficiently violated our assumptions of a balanced circuit so that our global depolarizing parameter estimation process is too unfaithful for adequate correction. We also see that without circuit balancing, the cost function (Equation~\ref{eqn:optimization}) increases monotonically, and does so much more quickly than with circuit balancing. We take this as evidence that the circuit balancing cost function is a good heuristic for whether a circuit is appropriate for this mitigation approach, since the optimized circuit balancing cost function is strongly correlated with mean Hellinger distance. 

\subsubsection{Depolarizing and Coherent Error}
\begin{figure}[]
    \centering
    \includegraphics[width=1.0\linewidth]{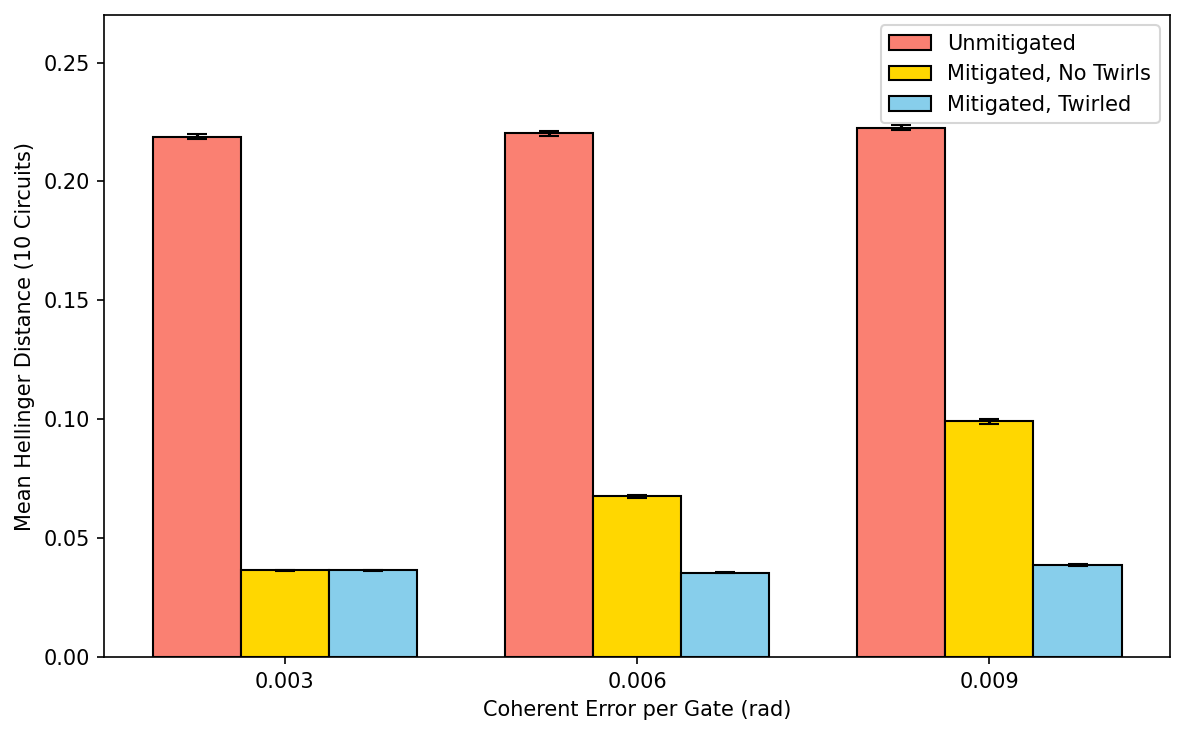}
    \caption{With varying amounts of coherent error introduced into the $n=10$ system in the form of XX rotations (x axis) in addition to depolarizing noise per gate drawn from a uniform distribution with maximum error $0.001$, we plot the mean Hellinger distance from ideal distributions (y axis) across unmitigated, untwirled mitigation, and twirled mitigation approaches. For twirled mitigation, we vary $t$ quadratically with the error strength (specifically, $t \in \{1, 4, 9\}$. The proportionality constant between twirls and squared error is thus $1/0.003^2=1.1\cdot10^5$ twirls/$\text{rad}^2$. We see that with appropriate twirling, our proposed error mitigation scheme significantly reduces the overall error, even in the presence of large coherent errors.}
    \label{fig:coherent_error}
\end{figure}

We now also introduce coherent error to the system with register size $n=10$ (once again fixing the depolarizing error bound to $0.001$ as in \ref{sec:scaling_size_results}) and run the protocol again to obtain the results for mean circuit Hellinger distance presented in Figure \ref{fig:coherent_error}.

First, we note that without mitigation, the mean Hellinger distance is quite high and is not significantly affected by varying the coherent error strength. 
This behavior is expected for systems with already high mixing, since the gate-level coherent error can be treated as another two-qubit gate within the circuit. 
We do not expect the presence of additional two-qubit gates to resist the effects of depolarizing mixing in biasing output distributions to become flatter over possible bistrings.

Next, we recall that the number of twirls, $t$, is varied quadratically, such that $t\in \{1,4,9\}$ corresponds to $\theta \in \{3\cdot 10^{-3},6\cdot 10^{-3},9\cdot 10^{-3}\}$. At $t=1$, corresponding to $\theta=0.003$, since there is only one twirling seed, this is identical to the untwirled setting. As a result, we expect identical performance between the untwirled and twirled methods when applied to mitigation for this value of $\theta$. 

However, for larger coherent error strengths, \textit{i.e.}, $\theta > 0.003$, despite mitigation the results worsen without additional twirling, and the untwirled Hellinger distance increases (yellow columns). However, as anticipated, increasing the number of twirls quadratically in coherent error rotation strength (blue columns) keeps the Hellinger distance almost constant across the tested rotation angles.

Lastly, to make our scaling arguments more concrete for practical use, we point out that at $\theta=0.009$ radians, two-qubit depolarizing parameters uniformly distributed over $[0.999,1]$, and a register size of $n=10$, we require only nine twirls and no two-qubit gate overhead to achieve an $83\%$ reduction in mean Hellinger distance over our $10$ test circuits. 
We note that the small overhead of nine twirls for these results over an ensemble of $10$ circuits with an average of $1983$ noisy CNOT gates would also be necessary for ZNE due to insensitivity to coherent error \cite{znebestpractices2023}. 
Additionally, executing circuit folding motifs for different noise factors would incur at least two times the total runtime to account for decay constant estimation, without even accounting for the time overhead associated with the circuit depth from folding \cite{znebestpractices2023}.

\section{Hardware Experiments \label{sec:hardware_experiments}}
\subsection{Methods \label{sec:hardware_methods}}
\subsubsection{Loschmidt Echo}
Our numerical experiments shed light on the capability of estimated depolarizing noise inversion under many idealized assumptions, including exact two-qubit gate error estimates, no shot noise, and no crosstalk. We now turn to the realistic setting of benchmarking this method on an actual quantum computer. In order to do this in the presence of shot noise, where it is exponentially difficult to precisely resolve bitstring probabilities, we turn to the Loschmidt echo \cite{loschmidt}. 
In this setting, we effectively construct a target approximate unitary 2-design, $U$, compose this with its inverse, and then apply $UU^\dagger$ to the all-zero computational basis state on a quantum computer. In the absence of error, we would expect the survival probability of the ground state to be exactly $1.0$, as we execute a resolution of the identity. We note that although $UU^\dagger$ resolves the identity, our depolarizing noise estimation process holds for any unitary $k$-design; as a result, we can approximate the depolarizing parameter for the sequence of gates implementing $U$ and $U^\dagger$. As such, our methods apply to the setting of a Loschmidt echo whose ``forward'' unitary, $U$ is an approximate 2-design.

\subsubsection{Circuit Construction}
We construct $10$ approximate unitary 2-designs, each over a $10$-qubit register, by implementing a forward unitary $U$ composed of $30$ layers. Each layer consists of a random single-qubit unitary applied to each register qubit, followed by a CNOT gate applied to each nearest-neighbor pair of logical qubits envisioning them on a linear chain. These logical circuits are then composed with an uncomputing sequence to implement $10$ Loschmidt echoes, each over $10$ qubits. The linear topology used in logical circuit construction makes this experiment amenable to target devices of many types of topologies, as compilation will produce no SWAP overhead as long as a linear chain of $10$ qubits exists in the device topology.

\subsubsection{Device Two-Qubit Depolarizing Parameter Extraction}
Our target quantum computer is IBM Fez, a superconducting qubit lattice with 156 qubits whose performance is regularly (daily) benchmarked and whose statistics are reported through IBM Quantum Cloud and can be accessed via Qiskit. 
Rather than directly reporting depolarizing parameters, this reports estimated gate infidelities from benchmarking, \textit{i.e.}, the 2-qubit gate infidelities $r$.  These can be related to the two-qubit gate depolarizing parameters $\lambda_2$ by 
\begin{equation}\label{eq:depoltoinf}
    r=\frac{3}{4}(1-\lambda_2),
\end{equation}
as is well-known \cite{irb2012}.
We can thus extract the two-qubit gate depolarizing parameter for each reported gate of IBM Fez.

\subsubsection{Compilation \& Circuit Balancing}
As mentioned previously, exactly solving the circuit balancing optimization problem is classically hard over a large device size. Instead, we use a Qiskit compiler pass manager with optimization level $2$ using SABRE routing and layout selecting methods to produce $100$ layouts, each from a different random seed \cite{sabre2019}. (We note that optimization level $3$ is incompatible with our goals in this section because it deterministically produces a fixed layout.) We then evaluate the circuit balancing cost function on each layout using the estimated gate depolarizations and choose the minimizing layout. For each of our $10$ circuits, we will refer to this minimal cost circuit as the ``balanced'' layout. We also choose the layout associated with the first random seed and refer to this as the ``default'' layout.

\subsubsection{Other Error Mitigation \& Device Run Parameters}
As mentioned in Section \ref{sec:practical_considerations}, we envision that the methods presented in this work will be used in conjunction with other error mitigation techniques. For our device runs, we have used $10000$ shots per circuit and $20$ Pauli twirls such that the shots are evenly distributed over each twirl. We also add measurement error mitigation in the form of Twirled Readout Error Extinction (TREX \cite{trex}) and dynamical decoupling. We expect the combination of these methods make the circuits more resilient to coherent error, dephasing and idling error, readout error, and two-qubit depolarizing error. However, we still expect to see the impacts of unmitigated crosstalk and single-qubit gate error. 

\subsubsection{Limitations of the Loschmidt Echo \label{sec:loschmidt_limitations}}
As mentioned, the Loschmidt echo produces a relevant experimental setting due to its resistance to shot noise in resolving arbitrary benchmarking distributions. However, when estimating circuit-wide depolarization, it will not penalize overestimates. In particular, if the estimation protocol described in this work significantly overestimates the circuit-wide depolarization induced by two-qubit gates when there are other sources of depolarization present, the Loschmidt echo test will erroneously report a better fidelity after inversion. This is because the method has erroneously cancelled some of the depolarization from other sources, for example, from single-qubit gates. Thus, we should only use the Loschmidt survival probability when we anticipate a good estimate of depolarization arising from two-qubit gates. As expected from the formalism and as shown in the simulations in Figure \ref{fig:variance_test}, good estimates for depolarization arising from two-qubit gates are produced for small circuit balancing cost function outputs, whereas large cost function outputs indicate aphysical parameter estimates (including a regime where inversion is more damaging than the unmitigated results). As a result, a successful run of our error mitigation strategy would imply a large Loschmidt echo survival probability and compiled circuits having layouts with a small circuit balancing cost function.

\subsection{Results \& Analysis \label{sec:hardware_results}}
\begin{figure}
    \centering
    \includegraphics[width=0.95\linewidth]{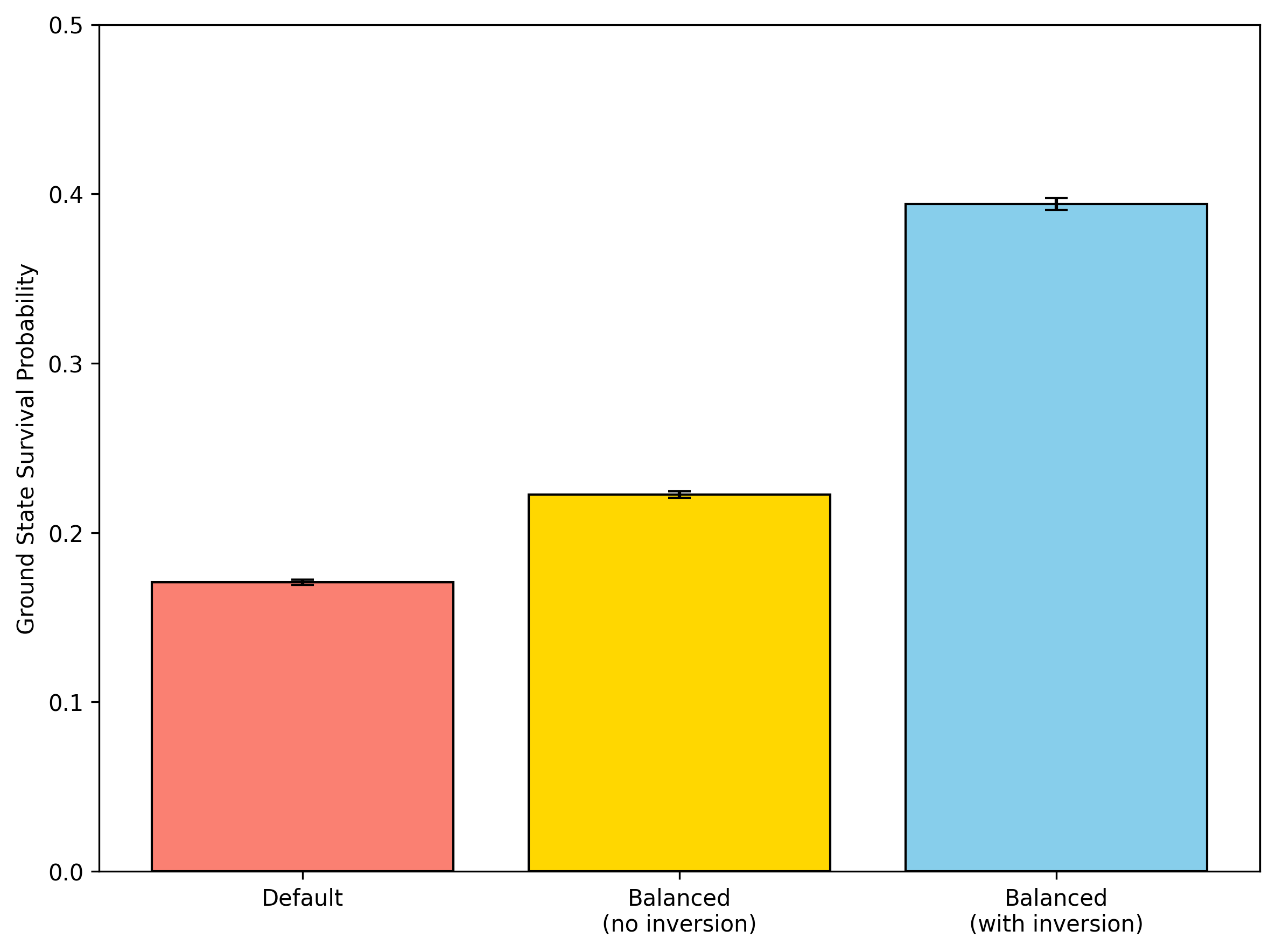}
    \caption{On IBM Fez, we find that our depolarizing parameter estimation and inversion scheme (enabled by circuit balancing) doubles the Loschmidt echo survival probability (y axis), a proxy for fidelity. The mitigation scheme here employed depolarizing parameters $\lambda_2$ extracted from estimated device infidelities $r$, as given by Equation \ref{eq:depoltoinf}.}
    \label{fig:device_run}
\end{figure}

The results for the experiment described in Section \ref{sec:hardware_methods} are summarized in Figure \ref{fig:device_run}. 
This shows that using circuit balancing and depolarization estimation with inversion as outlined in this work (blue column), we achieve more than double the ground state survival probability compared to the default layout with no inversion (red column). Additionally, we report a circuit balancing cost function value of $6.17\cdot10^{-5}$, a significant improvement over the default layout, which has a cost function value of $1.1\cdot 10^{-2}$. 
As a result, we do not display the result of the default layout with noise inversion in Figure \ref{fig:device_run}, since it produces an overinflated survival probability corresponding to an aphysical depolarization estimate, as described in Section \ref{sec:loschmidt_limitations} above. 
We also note that the balanced circuit with no inversion outperforms the default layout; we do not expect this to be generally true, since a balanced circuit may have larger depolarization effects for individual gates, even though the variance among expected local errors is small.

Overall, we consider the balanced run with noise inversion to be successful, as it provides a significant Loschmidt echo survival probability boost over the default while maintaining a small circuit balancing cost function. 
As anticipated, there is still significant error, potentially due to unmitigated crosstalk, single-qubit error, and relaxation effects. 
However, the error reduction is nonetheless significant and physically meaningful for relevant applications. The Loschmidt echo test corresponds to evaluating diagonal overlap matrix elements, for example, in a Gram matrix for quantum machine learning (QML) applications \cite{quantumkernels}. 
As such, we expect that for deep, highly entangling feature maps that approach unitary 2-designs, our error mitigation methods would produce much higher fidelity Gram matrices, leading to more precise bias-variance control \cite{heyraud2022}.
\subsection{Practical Considerations\label{sec:practical_considerations}}
Before concluding, we offer a few remarks on practical considerations of the approach presented here. First, we must assume a set of input circuits and device error distribution such that an approximate solution to the circuit balancing problem can be efficiently found. In practice, the objective for circuit balancing after a limited amount of optimization may be used as a heuristic for when a particular problem setting reasonably satisfies the assumptions made here. Section~\ref{sec:results} presented empirical evidence for the utility of this approach.

Next, the circuit balancing protocol assumes accurate gate-level benchmarking, as was emphasized in Section~\ref{sec:numerical_experiments}. For quantum devices with all-to-all connectivity, the number of benchmarking experiments would thus nominally scale quadratically in the number of qubits, which may make benchmarking itself unfeasible. However, in real-world instantiations of trapped-ion quantum computers, due to the identical nature of the ions, we would expect that the two-qubit gate error rate is largely defined by the gating zone to which a qubit is assigned \cite{qccd2021,h22023}. As a result, benchmarking the error per zone rather than per gate can produce a good estimate of gate error without incurring infeasible benchmarking times.

We further note that neglecting the potential additional depolarization arising from idling and single-qubit gates will produce an inaccurate estimate of the global depolarizing parameter. However, neglecting additional sources of depolarization will lead to an underestimate of Pauli support attenuation (and correspondingly, bitstring probability attenuation in the large register limit), due to the commutativity of depolarizing errors. As such, even if other depolarizing error sources are non-negligible, we can at least suppress the depolarization introduced by the two-qubit gates and produce a higher-fidelity output bitstring distribution compared to that obtained with no mitigation. Thus, the current approach is a worthwhile pursuit in most settings.

We have also implicitly assumed perfect knowledge of gate-level depolarizing error via a benchmarking protocol, but in practice this protocol may be imperfect. For example, assume that all two-qubit gates have the same depolarizing parameter, $\lambda$, but via benchmarking, we estimate $\hat{\lambda}=(1+\epsilon)\lambda$. Then, our estimate for the global depolarizing parameter for a balanced circuit will be $((1+\epsilon)\lambda)^{m\frac{15}{16}}$, where $m$ is the number of two-qubit gates. In the large register limit, our estimate for the noiseless setting will be $((1+\epsilon)\lambda)^{-m\frac{15}{16}}p_{\text{obs}}$, where $p_{\text{obs}}$ represents the observed probability. As a result, our method will yield a global depolarizing estimate that is off by a multiplicative factor of $(1+\epsilon)^{-m\frac{15}{16}}$. To reduce the negative impacts of imperfect gate error estimation, one can suppress the error bar of a randomized benchmarking protocol by increasing the number of sequences and/or shots per sequence until $\epsilon$ is sufficiently reduced to make the multiplicative error tolerable for target circuit gate counts.

Another source of error associated with randomized benchmarking is the presence of crosstalk. While it has been shown that idling crosstalk can be mitigated in certain settings \cite{qctrldd2024}, gate-based crosstalk can be present in a wide-variety of quantum computing architectures \cite{crosstalk2020}. As a result, there can be differences between errors benchmarked for isolated gates, compared with errors benchmarked for multiple gates in parallel.  
In this work, we have assumed that the strength of two-qubit depolarizing error is much greater than these spreading effects, (usually referred to as crosstalk). 
A simple device test such as parallel benchmarking sets of gates versus benchmarking gates in isolation can determine whether correlation effects are important.
If the benchmarked error is significantly different between these two settings, the mitigation scheme presented here may fail. There are also efficient methods to detect crosstalk rates that may serve as another heuristic to determining whether our mitigation procedure is appropriate for the particular target hardware \cite{crosstalk2020,simultaneousrb}. Additionally, depending on the nature of the target circuit, Simultaneous Randomized Benchmarking can enable error rate estimates under crosstalk \cite{simultaneousrb}. We anticipate that this may be a desirable method if the target circuit admits structured sequences of gate layers executed in parallel with nontrivial gate-based crosstalk.

Finally, we envision that the scheme presented here could be used in conjunction with other error mitigation schemes that seek to reduce effects, such as SPAM and idling noise manifesting in stochastic Pauli error \cite{qctrlsuppression2023}.

\section{Conclusions \& Future Work \label{sec:conclusions}} In summary, we have shown how two-qubit depolarizing elements can be approximated as functioning register-wide with a modified decay parameter when applied to balanced circuits representing unitary $k$-designs. Additionally, we have introduced a technique to enforce circuit balancing, enabling us to estimate this register-wide depolarizing parameter based on two-qubit gate depolarizing parameters, which are often publicly available through manufacturer benchmarking efforts. We then illustrate that using the estimated circuit-wide global depolarizing parameter, we can reduce depolarizing noise by first suppressing coherent error and then directly inverting the survival function. We constructed an asymptotic scaling estimate on the number of twirls needed to satisfy a certain error condition on resulting bitstring distributions. 
We then showed with numerical simulations that the methods presented here produce significant reductions in Hellinger distance, and are robust to increases in register sizes, coherent error strengths, and asymmetries in circuit design or device error distribution. Lastly, we demonstrated that our methods produce tangible performance improvements when applied to a real quantum device using the Loschmidt echo test. 

We present this technique as a method to avoid costly gate or circuit folding techniques that increase circuit depth and consequently wall-clock time. Crucially, our method adds no two-qubit gate overhead and does not rely on low-entanglement regimes for tractability. We anticipate maximal real-world utility for simulations of chaotic quantum dynamics or any circuit-based application approaching a unitary $k$-design, when applied to near-term and early fault tolerant quantum computers.

We also provide a few comments here on the utility and advantage of this approach to quantum circuit error mitigation. It is well known that asymptotically for large circuit depths under constant depolarizing noise per gate, there exists a classical polynomial-time algorithm to approximate noisy random circuit sampling \cite{polynomialnoisesim}. Further, it has recently been shown that any geometrically local quantum circuit admits polynomial-time approximate classical sampling beyond a depth threshold that is related to the magnitude of circuit noise \cite{geometricallylocal}. However, there exists a finite depth regime where these results do not hold, and the methods presented in this work have empirically demonstrated success in such a regime of intermediate depth, as indicated by the hardware results in Section~\ref{sec:hardware_results}. Consequently, we envision that an error mitigation scheme such as the one presented in the current work would have utility for intermediate-depth applications that exhibit quantum chaos, potentially beyond the possibility of classical simulation.

There are several opportunities for future work, the most prominent of which is to factor in single-qubit gate error into the global depolarization parameter estimate. We envision this may be done by formulating an optimization problem with two objectives: one for two-qubit gate placement and one for single-qubit gate placement, analogous to our circuit balancing technique. However, the weights of each objective may require heuristic tuning. Second, we leave shot noise propagation to future work; in regimes with large depolarization, we anticipate a nontrivial shot overhead arising from resolving small probability differences in bitstring distributions. In our work, we have implicitly assumed depolarization regimes small enough to avoid this issue. Lastly, further, more extensive testing on larger systems is left to future work.\newline

\section{Acknowledgments}
The authors are grateful for productive conversations with and recommendations from Jordan Hines, technical suggestions from Aswath Suryanarayanan, and feedback from Yilun Yang. 

This work was supported by the Defense Advanced Research Projects Agency under Grant Number HR0011- 24-9-0358 and by the U.S. Department of Energy, Office of Science, National Quantum Information Science Research Centers, Quantum Systems Accelerator.

\bibliographystyle{unsrt}
\bibliography{refs}
\end{document}